\documentclass[a4paper,11pt]{article}
\usepackage{graphicx}
\usepackage{lipsum}
\usepackage{url}
\usepackage{float}
\usepackage[utf8]{inputenc}
\usepackage{orcidlink}
\usepackage{amsmath}
\usepackage{graphicx,amsfonts,amssymb,slashed}
\usepackage{bbold,wasysym}
\usepackage{enumitem}
\usepackage{booktabs}
\usepackage{bbm,euscript,hyperref}
\hypersetup{colorlinks, citecolor=blue, linkcolor=red, urlcolor=blue}
\usepackage{color}
\usepackage[numbers, sort&compress]{natbib}
\usepackage{etoolbox}
\usepackage{hyperref}
\usepackage{tikz}
\usepackage{orcidlink} % Must come after hyperref and tikz
\title{A Stochastic Approach for Determining the Quark Confinement Potential of Charmonia}

\author{
Ahmet Bingül$^{1,2}$\thanks{
\href{https://orcid.org/0000-0002-2455-8039}{ORCID: 0000-0002-2455-8039},
bingul@gantep.edu.tr}
\and
Altuğ Özpineci$^{2}$\thanks{
\href{https://orcid.org/0000-0003-0791-7658}{ORCID: 0000-0003-0791-7658},
ozpineci@metu.edu.tr}
}

\date{
$^1$Department of Engineering of Physics, Gaziantep University, Gaziantep 27310, Türkiye\\
$^2$ Department of Physics, Middle East Technical University, Ankara 06800, Türkiye
}
\begin{document}

\maketitle

\begin{abstract}
In this study, a non-relativistic potential model is used to calculate the mass spectrum and decay properties of low lying charmonium states. A stochastic framework is proposed to extract the possible analytical form of the confinement part of the interaction potential between quarks. Based on this approach, it is found that the confinement function deviates slightly from a linear form at large distances. The results obtained are compared with other theoretical predictions and current PDG values. The obtained results are also applied to the special case of $X(3872)$, and the ratio of its radiative decays is obtained as $R_{\psi\gamma}=2.10$ which is about $2\sigma$ larger than the experimental result.
This can be interpreted as the presence of a  significant molecular contribution in $X(3872)$.
\end{abstract}

%
% the environments 'definition', 'lemma', 'proposition', 'corollary',
% 'remark', and 'example' are defined in the LLNCS documentclass as well.
%

\maketitle
\noindent

%%%%%%%%%%%%%%%%%%%%%%%%%%%%%%%%%%%%%%%%%%%%%%%%%%%%%%%%%%%%%%%%%%%%%%%%%%%%%%
\section{Introduction}
\label{sec:intro}

According to Quantum Chromodynamics (QCD), conventional mesons are bound states of quarks and anti-quarks ($q \bar q$), such as $\pi^+(u \bar d)$. The quark model explained the Eightfold Way and correctly predicted the structure of the proton successfully as well.
With the discovery of $J/\psi(1S)$ and $\psi(2S)$ mesons in 1974, hadron spectroscopy became a remarkable area for testing and exploring the properties of the strong interactions predicted by QCD. The charmonium system, bound states of $c \bar c$ quarks, allows the prediction of some of the parameters of the states using non-relativistic~\cite{Cui, Kher, Soni} and relativistic potential models~\cite{GI, Barnes}, lattice QCD~\cite{LQCD}, and sum rules~\cite{Cho, Brambilla}. 

On the other hand, when attempting to explain certain properties of some observed mesons within simple $q \bar q$ picture, various difficulties arise. Therefore, it is suspected that these newly observed mesons may correspond to more complex configurations than simple $q \bar q$ pairs. One of these exotic states is the $X(3872)$ meson. 
Since its mass lies close to the $D^0$-$\bar D^{*0}$ threshold, it is considered that it might be a molecular state composed of $D$ and $D^*$ mesons \cite{Gamermann:2009uq} or it may have more complex exotic form \cite{Cincioglu:2016fkm}. Hence, more experimental observations are required to understand the structure of such states. The radiative decay widths of $X(3872)$ state are used to probe  its nature. These decays have been extensively studied within theoretical frameworks (see e.g. \cite{Colangelo:2025uhs,Guo:2014taa, Cincioglu:2016fkm}). A  recent experimental result obtained by the LHCb Collaboration~\cite{LHCbratio} for the radiative decays of $X(3872)$ can provide valuable information to these theoretical models. The list of observed $q\bar{q}$ mesons and possible non-$q\bar{q}$ states is provided by the Particle Data Group (PDG)~\cite{PDG}.

In this article, a non-relativistic model is used to calculate the mass spectrum and decay properties of several charmonium states. The interaction potential is generally modeled as a sum of two terms; the coulomb repulsive part and quark confinement potential. For the confinement part, various forms are tested in the literature such as linear, exponential or logarithmic functions.
In this study, we propose a stochastic approach to \textit{derive} the possible analytical form of the confinement function. In short, first, a stochastic function is constructed by a collection of random sample points in increasing order to account for charmonium mass spectra, then an appropriate function is fitted to these points to obtain final shape of the confinement component of the potential.

Section~\ref{sec:method} describes the theoretical framework and construction of the confinement function. The decay properties and radiative transitions for charmonium states are given in Section~\ref{sec:decayproperties}. Radiative decays of the $X(3872)$ meson are briefly discussed in Section~\ref{sec:x3872}. Section~\ref{sec:theend} presents the summary and conclusion of this study. Furthermore, to demonstrate the capabilities and limits of the method on saturating potentials, an inverse reconstruction benchmark using synthetic mock data is provided in the Appendix.

%%%%%%%%%%%%%%%%%%%%%%%%%%%%%%%%%%%%%%%%%%%%%%%%%%%%%%%%%%%%%%%%%%%%%%%%%%%%%%
\section{Method}
\label{sec:method}

\subsection{Dynamics of Charmonium}
The $c \bar c$ system can be considered as a non-relativistic bound state, since the quark masses are much larger than their kinetic energies. Consequently, solving the Schrödinger equation allows us to determine the mass spectrum and associated wavefunctions of the charmonium states.

Nonrelativistic Hamiltonian that governs the bound state $c \bar c$ system can usually be written as 
\begin{equation}
\label{eqn:hamiltonian}
  H = \frac{p^2}{2m_r} + V(r) + 2m_c  
\end{equation}
where $p$ is the relative momentum of each quark, $m_c$ is the charm quark (or antiquark) mass, $m_r = m_c/2$ is the reduced mass of the system. $V(r)$ is the spherically symmetric interaction potential energy function and $r$ is the interquark separation. For a given angular momentum quantum number, $L \ge 0$, the radial part of the time-independent Schrödinger Equation in natural units is given by
\begin{equation}
\label{eqn:radial}
  \left[-\frac{1}{2m_rr^2} \frac{d}{dr} \left(r^2\frac{d}{dr}\right) + \frac{L(L+1)}{2 m_r r^2} + V(r) \right] R(r) = ER(r) 
\end{equation}
where $E$ is the total energy of the $c \bar c$ system. By defining $u(r) = rR(r)$, we obtain a reduced form of the Schrödinger Equation as follows
\begin{equation}
\label{eqn:reduced}
  \left[ -\frac{1}{2m_r} \frac{d^2}{dr^2} + \frac{L(L+1)}{2m_rr^2} + V(r) \right] u(r) = Eu(r) 
\end{equation}
The form of $V(r)$ is generally nontrivial. Therefore, numerical methods are implemented to solve Equation~\ref{eqn:reduced}. In this study, we employ the Numerov Method for Equation~\ref{eqn:reduced} to determine the charmonium eigenvalues and wavefunctions.

For the bound $c \bar c$ system, the interaction potential may be decomposed as
\begin{equation}
\label{eqn:pot}
  V(r) = -\frac{4}{3}\frac{\alpha_s}{r} + V_{co}(r) + V_{sd}(r)
\end{equation}
Here, the first term corresponds to a Coulomb-like interaction describing the one-gluon exchange between quarks and it is dominant at short interquark distances. $\alpha_s$ is a strong running coupling constant. 

$V_{co}(r)$ is the confinement component of the interaction and is expected to be dominant at large distances. In 1970s, it was proposed as a linear function $V_{co}(r) = br$ to account for the observed masses of quarkonium states where $b$ is the QCD tension, corresponding to the energy per unit length of the flux tube formed by the collapsing gluonic field lines\footnote{The sum of first term and linear confinement function is known as the Cornell Potential. At first, $\alpha_s$ and $b$ were empirical model parameters. However, with the development of QCD, they are calculated via perturbative QCD and lattice QCD.}. The linear function  has provided good agreement with the experimental data for quarkonium mass spectroscopy~\cite{Kher, Soni}. 

On the other hand, several alternative forms of the confinement potential have been proposed in the literature, 
such as the screening function $V_{co}(r) = b(1-\exp(-r/\mu))/\mu$~\cite{Cui}, logarithmic function $V_{co}(r) = b\log(r/\mu)$~\cite{Logarithmic},  and the power function $V_{co}(r) = b r^\nu$~\cite{PowerFunc1, PowerFunc2} where $b$, $\mu$ and $\nu$ are free numerical parameters. The results of these studies show that these alternative forms are also in good agreement with PDG. Consequently, it is quite natural to ask what the precise shape of the confinement function is. To answer this, we introduce a {\it stochastic} way to explore its possible form, the details of which are given in the next section.

The last term in the potential accounts for the spin-dependent interactions, which are essential for removing the degeneracy of the charmonium states. This term is a function of total angular momentum, $\mathbf{J=L+S}$, and total spin, $\mathbf{S=S_c + S_{\bar c}}$ where $\mathbf{S_{c, \bar c}}$ stands for the spin of individual quarks. Hence, $S = 0$ or $1$ (corresponding to singlet or triplet states). The spin dependent-term is generally expressed as~\cite{Kher, Soni}
\begin{equation}
\label{eqn:spin}
\begin{split}
V_{sd}(r) ={} & V_{SS}(r) (\mathbf{S_c} \cdot \mathbf{S_{\bar c}}) +
                V_{LS}(r) (\mathbf{L} \cdot \mathbf{S})  +  \\
              & V_{T}(r) \left[ S(S+1) - 3 (\mathbf{S} \cdot \mathbf{\hat{r}}) (\mathbf{S} \cdot \mathbf{\hat{r}}) \right]
\end{split}
\end{equation}
where 
$\left <\mathbf{S_c} \cdot \mathbf{S_{\bar c}} \right> = S(S+1) - 3/2$ and
$\left <\mathbf{L} \cdot \mathbf{S} \right> = [J(J+1)-L(L+1)-S(S+1)]/2$.
The spin-spin interactions can be written as
\begin{equation}
\label{eqn:vss}
V_{SS}(r) = \frac{1}{3m_Q^2} \nabla^2 V_V = \frac{16\pi\alpha_s}{9m_Q^2} \delta^3 (r)
\end{equation}
where $\delta^3(r) = (\sigma/\sqrt{\pi})^3 \exp(-\sigma^2 r^2)$ and $\sigma$ is a smearing constant \cite{GI}. Finally, the spin-orbit and tensor interactions can be written in terms of the vector and scalar parts of $V(r)$ as follows  \cite{Voloshin:2007dx}:
\begin{equation}
\label{eqn:vls}
V_{LS}(r) = \frac{1}{2m_Q^2 r} \left( 3\frac{dV_V}{dr} - \frac{dV_S}{dr} \right)
\end{equation}
\begin{equation}
\label{eqn:vte}
V_{T}(r) = \frac{1}{6m_Q^2} \left( 3\frac{d^2 V_V}{dr^2} - \frac{1}{r} \frac{dV_V}{dr} \right).
\end{equation}
where $V_V = -4\alpha_s/3r$ and $V_S = V_{co}(r)$.

Optimal values of the parameters ($\alpha_s, m_c, \sigma)$ and the parameters of $V_{co}(r)$ are obtained by fitting the resulting eigenvalues to the current PDG data, specifically the well-known $c \bar c$ states. In general, optimization is performed by minimizing a $\chi^2$ value of the following form
\begin{equation}
\label{eqn:chi2}
\chi^2 = \sum_k^{N_{c}} \left( \frac{W_k-M_k}{\varepsilon_k}  \right)^2
\end{equation}
where $W_k$ is the evaluated mass of $k^\text{th}$ state obtained from numerical solution, and $M_k$ and $\varepsilon_k$ are the experimental values of the observed mass and associated uncertainty (total intrinsic width), respectively. For this study, $N_c = 17$ charmonium states are used in the fitting procedure. The mass spectrum and decay calculations are restricted to states up to $n=4$ (below $\sim 4.5\text{ GeV}$) where non-relativistic potential approximations remain theoretically valid.~\cite{Barnes}. 

Charmonium bound states are categorized using the spectroscopic notation $n^{2S+1}L_J$. In this convention, $n=1,2,\dots$ represents the principal (radial) quantum number. As compiled in PDG data~\cite{PDG}, these states can also be classified by their $J^{PC}$ quantum numbers, where parity $P =(-1)^{L+1}$ and charge conjugation $C =(-1)^{L+S}$. Using this notations, the predicted charmonium masses are listed in Table~\ref{tab:fullmassspectra}, with values rounded to the nearest 1 MeV.  Using the potential parameters and numerical values of the radial wavefunctions, the decay constants, decay widths, and the transition matrix elements can be calculated as described in Section~\ref{sec:decayproperties}.

%\begin{figure}[htb]
%\centering
%\includegraphics[width=0.8\textwidth]{fig_1S0.pdf}
%\caption{(Color online) 
%Reduced $S$-state wavefunctions obtained using a linear confinement potential. The %wavefunctions are nearly identical for other confinement potentials.}
%\label{fig:1S0}
%\end{figure}

\begin{table}[hp] % The asterisk makes it span both columns
\centering
\small
\caption{Comparison of some predicted charmonia masses in MeV with other works and experimental data. The values in the parenthesis are the total width of the particle in PDG data. We use the confinement function given in Equation~\ref{eqn:soner}. Optimal potential parameters are selected as $(\alpha_s,\sigma,m_c, b)=(0.487, 1.162, 1.433,0.135)$}

\label{tab:fullmassspectra}
\begin{tabular}{ lcccccc }
\toprule
$n^{2S+1}L_J$ & $J^{PC}$    & This Work  & \cite{Cui} & \cite{Kher} & \cite{Soni} & PDG~\cite{PDG} \\
\midrule
$1 ^1S_0$     & $0^{-+}$    & 2994  & 3006   & 2995        & 2989        & 2984(31)   \\
$2 ^1S_0$     & $0^{-+}$    & 3644  & 3644   & 3606        & 3602        & 3638(12)   \\
$3 ^1S_0$     & $0^{-+}$    & 4039  & 4044   & 4000        & 4058        &            \\
$4 ^1S_0$     & $0^{-+}$    & 4348  & 4353   & 4328        & 4448        &            \\
\midrule
$1 ^3S_1$     & $1^{--}$    & 3097  & 3097   & 3094        & 3094        & 3097(0.1)  \\
$2 ^3S_1$     & $1^{--}$    & 3684  & 3686   & 3649        & 3681        & 3686(0.3)  \\
$3 ^3S_1$     & $1^{--}$    & 4066  & 4073   & 4036        & 4129        & 4040(84)   \\
$4 ^3S_1$     & $1^{--}$    & 4369  & 4375   & 4362        & 4514        & 4415(5)    \\
\midrule
$1 ^3P_0$     & $0^{++}$    & 3551  & 3405   & 3457        & 3428        & 3415(11)   \\
$2 ^3P_0$     & $0^{++}$    & 3959  & 3846   & 3866        & 3897        & 3922(20)   \\
$3 ^3P_0$     & $0^{++}$    & 4278  & 4186   & 4197        & 4296        &            \\
$4 ^3P_0$     & $0^{++}$    & 4548  & 4467   &             & 4653        &            \\
\midrule
$1 ^1P_1$     & $1^{+-}$    & 3524  & 3521   & 3534        & 3470        & 3525(1)    \\
$2 ^1P_1$     & $1^{+-}$    & 3934  & 3940   & 3936        & 3943        &            \\
$3 ^1P_1$     & $1^{+-}$    & 4254  & 4262   & 4269        & 4344        &            \\
$4 ^1P_1$     & $1^{+-}$    & 4525  & 4527   &             & 4704        &            \\
\midrule
$1 ^3P_1$     & $1^{++}$    & 3564  & 3515   & 3523        & 3468        & 3511(1)    \\
$2 ^3P_1$     & $1^{++}$    & 3968  & 3935   & 3925        & 3938        & 3872(1)    \\
$3 ^3P_1$     & $1^{++}$    & 4286  & 4258   & 4257        & 4338        & 4286(51)   \\
$4 ^3P_1$     & $1^{++}$    & 4554  & 4523   &             & 4704        &            \\
\midrule
$1 ^3P_2$     & $2^{++}$    & 3552  & 3539   & 3556        & 3480        & 3556(2)    \\
$2 ^3P_2$     & $2^{++}$    & 3959  & 3957   & 3956        & 3955        & 3923(35)   \\
$3 ^3P_2$     & $2^{++}$    & 4278  & 4278   & 4290        & 4358        &            \\
$4 ^3P_2$     & $2^{++}$    & 4547  & 4541   &             & 4718        &            \\
\midrule
$1 ^3D_1$     & $1^{--}$    & 3809  & 3800   & 3799        & 3755        & 3774(27)   \\
$2 ^3D_1$     & $1^{--}$    & 4150  & 4143   & 4145        & 4188        & 4191(5)    \\
$3 ^3D_1$     & $1^{--}$    & 4434  & 4423   & 4448        & 4555        &            \\
$4 ^3D_1$     & $1^{--}$    & 4682  & 4660   &             & 4891        &            \\
\midrule
$1 ^3D_2$     & $2^{--}$    & 3814  & 3808   & 3805        & 3772        & 3824(2)    \\
$2 ^3D_2$     & $2^{--}$    & 4152  & 4152   & 4152        & 4188        &            \\
$3 ^3D_2$     & $2^{--}$    & 4435  & 4432   & 4456        & 4557        &            \\
$4 ^3D_2$     & $2^{--}$    & 4682  & 3807   &             & 4896        &            \\
\bottomrule
\end{tabular}
\end{table}

%%%%%%%%%%%%%%%%%%%%%%%%%%%%%%%%%%%%%%%%%%%%%%%%%%%%%%%%%%%%%%%%%%%%%%%%%%%%%%
\subsection{Extraction of Confinement Function}
\label{sec:stochastic}
The common characteristics of the phenomenological confinement functions discussed in the previous section are that they are single-valued and monotonically increasing. Based on these mathematical facts, the shape of the confinement function can be approximately determined using a stochastic approach (or Monte Carlo method) as follows. 

Initially, the potential parameters $(\alpha_s, m_c, \sigma)$ are randomly selected within physically allowed ranges, e.g. $\alpha_s=[0.3, 0.6]$, $m_c=[1.2, 1.6]$ GeV and $\sigma=[0.5, 1.5]$ GeV. Then, a set of sample points $(\rho_j, V_j)$ of size $N$ is generated such that the radial values ($\rho_j$) are sampled uniformly and the stochastic confinement potential values ($V_j$) are selected randomly in increasing order, namely $V_{j+1} > V_j$ and $\rho_{j+1} > \rho_j$ for $j=0,1,\dots,N-1$. Starting with the initial values $\rho_0=V_0=0$, successive values are evaluated iteratively using the following steps
\begin{equation}
\label{eqn:slopes}
\begin{split}
  \rho_{j+1} &= \rho_j + \Delta \rho \\
  V_{j+1} &= V_j + c_j\Delta \rho
\end{split}
\end{equation}
where the step size\footnote{$\Delta \rho$ is much greater than the step size used in Numerov method.} is $\Delta \rho = (r_{max}-r_{min})/(N-1)$. 

\begin{figure}[htb]
\centering
\includegraphics[width=0.8\textwidth]{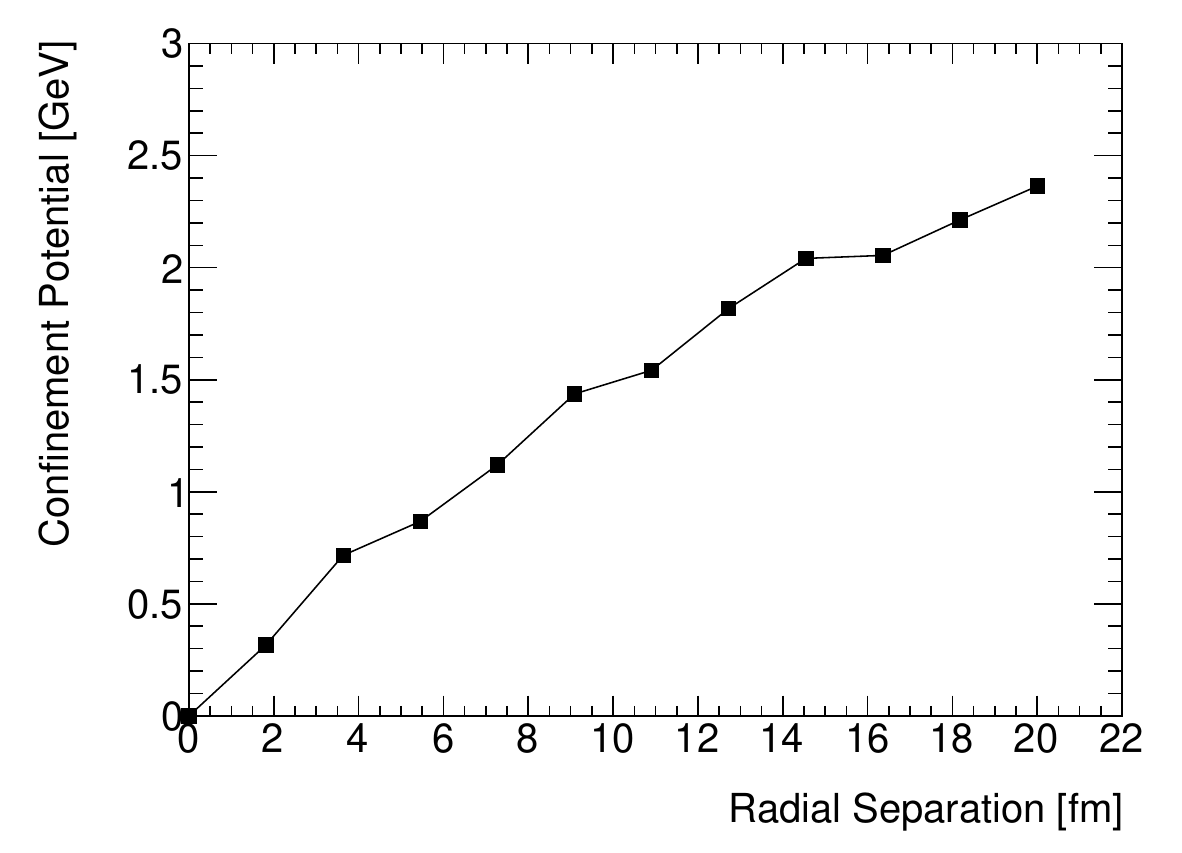}
\caption{An example stochastic confinement potential function having a large fluctuation for $N=12$ sample points obtained by a stochastic way described in text after 7 million trials for $\chi^2_{\text{max}} = 750$.}
\label{fig:stocastic}
\end{figure}

The stochastic slopes $c_j$, expressed in the corresponding energy-per-length units, are sampled uniformly from the interval $[0, 1/2]$\footnote{The upper limit of this interval significantly affects the computational time to obtain the $V_j$ values with less fluctuations.} across all radial distances. This domain provides an unconstrained parameter space containing both linear ($c_j \approx \text{const}$ for all points) and sub-linear ($c_j \to 0$) behaviors. 
A candidate configuration is randomly constructed according to Equation~\ref{eqn:slopes}. Values between two neighboring points are computed by linear interpolation. Subsequently, Equation~\ref{eqn:reduced} is solved to have eigenvalues and the $\chi^2$ value is evaluated for this configuration at the end. 
These steps are repeated until the candidate configuration satisfies a strict goodness-of-fit threshold of $\chi^2 \le \chi^2_{\text{max}}$ against $N_c = 17$ experimental charmonium states. A trial potential is accepted only when its evaluated mass spectrum meets this rejection sampling criterion.
Note that due to the high dimensionality of the random slope parameter space, several million iterations can be required to achieve a single valid configuration depending on $\chi_{\text{max}}^2$ value. 
Lastly, the optimal parameters $(\alpha_s, m_c, \sigma)$ and $V_j$ are preserved for further functional analysis. An example confinement potential for $N = 12$ points is shown in Figure~\ref{fig:stocastic}.

\begin{figure}[htbp]
\centering
\includegraphics[width=0.8\textwidth]{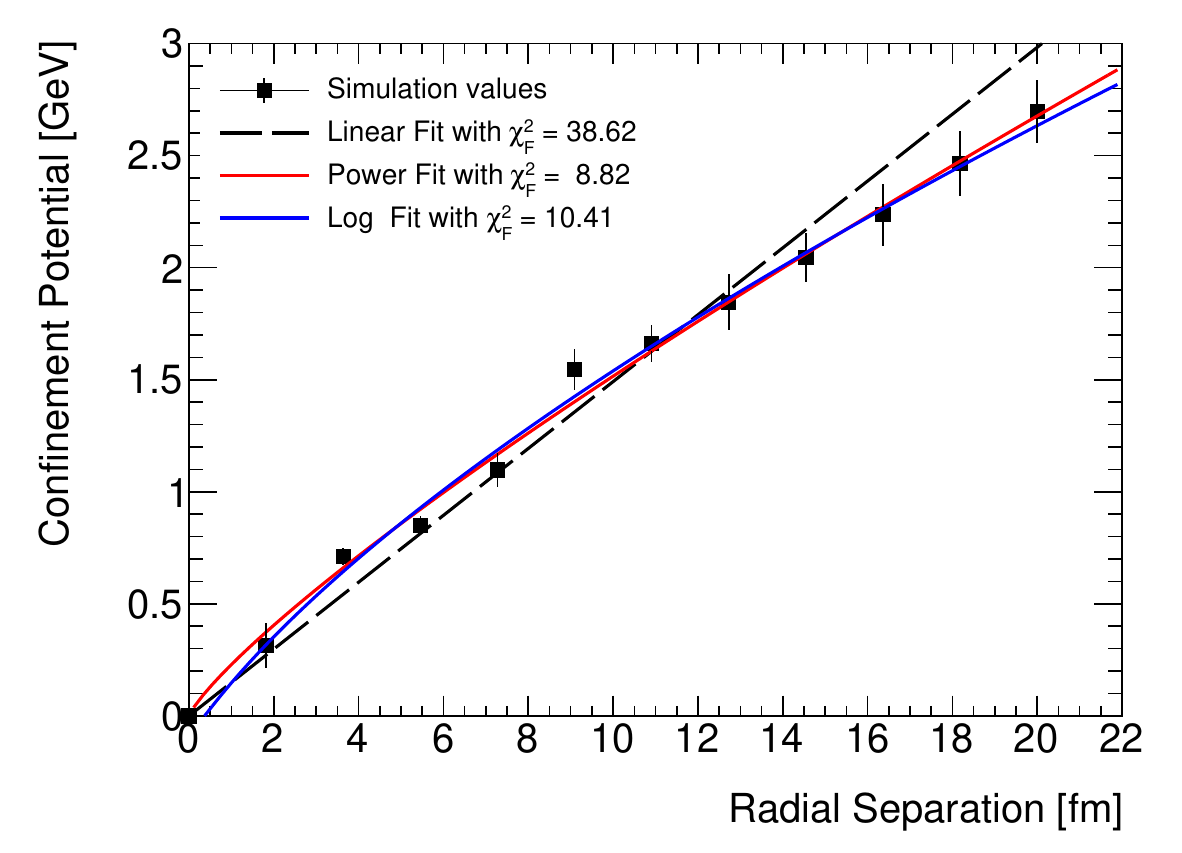}
\caption{(Color online) Average stochastic confinement potential values obtained from average of 20 independent configurations for $N=12$ points. Three different functions are fitted to the simulated data. Error bars represent one standard deviation.}
\label{fig:meanpot12}
\end{figure}

\begin{figure}[htbp]
\centering
\includegraphics[width=0.8\textwidth]{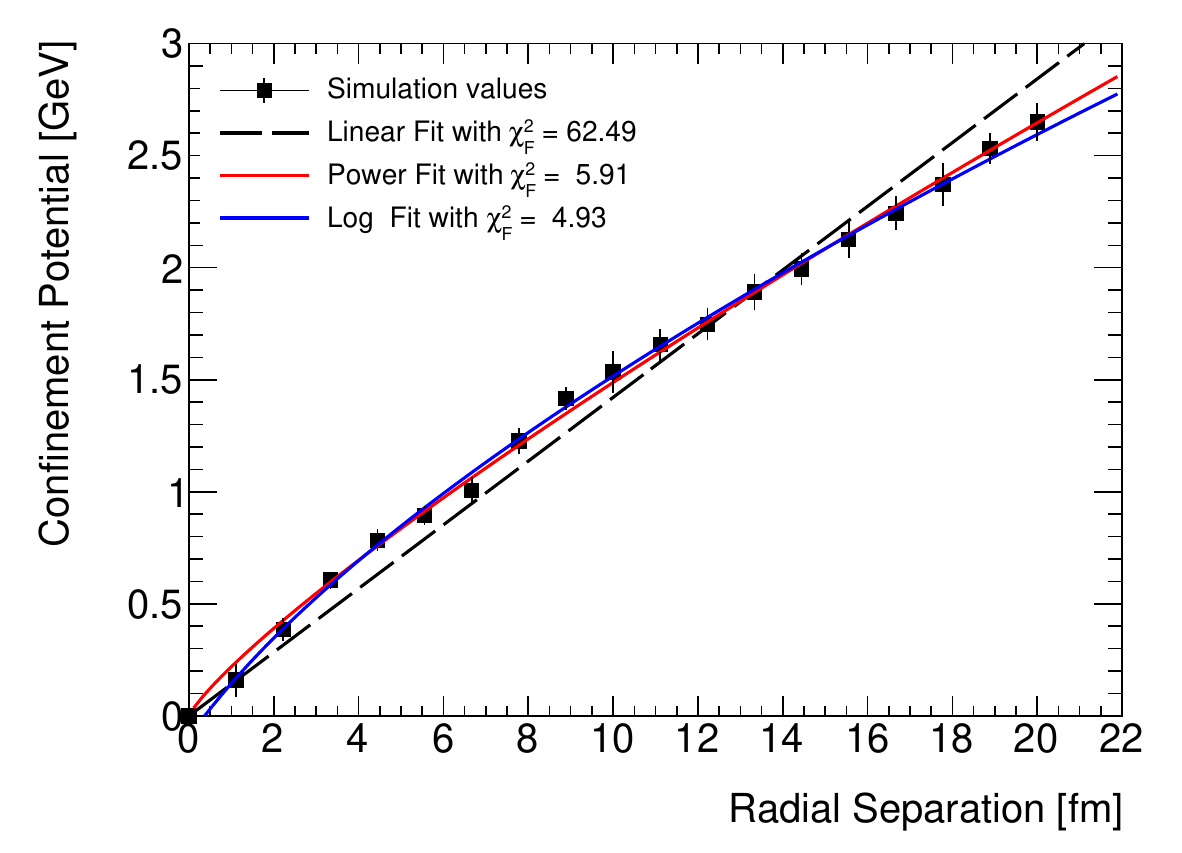}
\caption{(Color online) Average stochastic confinement potential values obtained from average of 20 independent configurations for $N=19$ points.}
\label{fig:meanpot19}
\end{figure}

 Because an individual accepted potential retains localized stochastic fluctuations inherent in single random seed realizations (Figure~\ref{fig:stocastic}), an ensemble averaging approach is employed. The algorithm is repeated using independent random seeds to generate $N_{config}$ valid configurations. We monitored the point-by-point statistical standard deviation $\sigma_j$ and standard error $\sigma_j / \sqrt{N_{config}}$. Ensemble stability and spatial smoothness\footnote{A smooth function typically exhibits a relatively small second derivative which can be computed numerically at each point (for $j = 1,2,\dots,N-2$).} saturate at $N_{config}=20$; further increasing the number of configurations yields no statistically significant shift in the extracted confinement potential (or fit parameters). Consequently, mean values and standard deviations from 20 independent configurations are used for subsequent functional fits. The resulting mean potentials, obtained by averaging over the 20 stochastic values are shown in Figure~\ref{fig:meanpot12} an Figure~\ref{fig:meanpot19} for $N = 12$ and $N = 19$ points, respectively.

% Obviously the algorithm does not force a uniform slopes at large distances and independently selects configurations where the effective slopes $c_j$ slightly tend to decrease at larger interquark separations. This ensures that sub-linear confinement shape (as in the logarithmic, power and screening [string-breaking] functions) is fairly sampled without a priori algorithmic bias.

It is important to note that the stochastic procedure does not assign equal probability to different functional forms. Although the independent sampling of ($c_j$) permits local slopes to decrease and therefore allows sub-linear configurations to occur, extended regions with very small slopes are statistically suppressed by the proposal distribution. Consequently, the method can explore deviations from linear confinement within the sampled parameter space, but it should not be interpreted as an unbiased reconstruction of all possible confinement functions. The ability of the method to recover strongly screened potentials is therefore tested separately using synthetic benchmark data in the Appendix.


Finally, to obtain the analytical form of the potential, the sample points are fitted to linear ($br$), 
power ($b r^\nu$), and a logarithmic function of the following form:
\begin{equation}
\label{eqn:soner}
    V_{\text{co}}(r) = b \sqrt{\frac{r}{\mu}} \left(1 + \ln\left(\frac{r}{\mu}\right)\right)
\end{equation}
 Here, $\mu$ is introduced to ensure a dimensionless argument in the logarithm. When $\mu$ was initially varied as a free parameter in trial fits, it consistently converged near $1.0\text{ fm}$. Thus, fixing $\mu = 1\text{ fm}$ reduces the confinement model to a single free scaling parameter $b$. The fit results and corresponding $\chi_F^2$ values arising from the fitting process are shown in the same figures.

 In Ref~\cite{M:2023hms}, various potential models used in the literature are presented. One of them is the Song and Lin potential (in our case, we have the coefficient of the $r^{-1/2}$ term is set to zero). A logarithmic correction can be physically motivated by observing that in higher order QFT computations, the corrections are typically logarithmic. Although in our case the potential is not a result of a higher order QFT computation, it would not be surprising to have logarithmic dependence in the potential.

%%%%%%%%%%%%%%%%%%%%%%%%%%%%%%%%%%%%%%%%%%%%%%%%%%%%%%%%%%%%%%%%%%%%%%%%%%%%%%
\section{Decay Properties and Transitions}
\label{sec:decayproperties}
After determining the confinement potential function, Equation~\ref{eqn:soner}, we optimize the $\chi^2$ function again to obtain the final form of the parameters as $(\alpha_s,\sigma,m_c, b)=(0.487, 1.162, 1.433, 0.135)$.
Using these parameters, the eigenvalues and the wavefunctions are achieved to evaluate the decay properties of the charmonium states.

\subsection{Leptonic Decay Constants}
In nonrelativistic limit it is common to use the Van-Royen-Weisskopf formula~\cite{Royen} to calculate $c \bar c$ leptonic decay constants of pseudoscalar and vector mesons with QCD correction factors as follows
\begin{equation}
\label{eqn:fp}
f_P^2 = \frac{3|R_{ns}(0)|^2}{\pi M_{nS}} \left(1-\frac{2\alpha_s}{\pi} \right)
\end{equation}
\begin{equation}
\label{eqn:fv}
f_V^2 = \frac{3|R_{ns}(0)|^2}{\pi M_{nV}} \left(1-\frac{8\alpha_s}{3\pi} \right)
\end{equation}
The computed and observed values of the decay constants are listed in Table~\ref{tab:decayconst}.
\begin{table}[htbp] % The asterisk makes it span both columns
\centering
\small
\caption{Pseudoscalar and vector meson decay constants in MeV.}
\label{tab:decayconst}
\begin{tabular}{ lcccccc }
\toprule
Decay Constant & State & This Work  & \cite{Kher} & \cite{Soni} & \cite{PDG}    \\
\midrule
$f_P$          & 1S   &  498  & 501        & 350         & $335 \pm 75$   \\
               & 2S   &  294  & 301        & 278         &                \\
               & 3S   &  237  & 264        & 249         &                \\
               & 4S   &  207  & 245        & 231         &                \\
\midrule
$f_V$
               & 1S   &  361  & 510        & 326         & $411 \pm 5$    \\
               & 2S   &  246  & 303        & 257         & $271 \pm 8$    \\
               & 3S   &  206  & 265        & 230         & $174 \pm 18$   \\
               & 4S   &  183  & 240        & 213         &                \\
\bottomrule
\end{tabular}
\end{table}

\subsection{Annihilation Decays}
The measurements of the two-body decay widths offer valuable information about dynamics of $c \bar c$ mesons.

The annihilation decay width of the charmonium states into two photons with leading order QCD corrections are given by Van Royen-Weiskopf formula~\cite{Royen}:
\begin{equation}
\Gamma(n^1 S_0 \to \gamma\gamma) = 
        \frac{3\alpha_e^2 Q^4 |R_{nS}(0)|^2}{m_c^2} 
        \left[1 -  \frac{3.4\alpha_s}{\pi}\right]
\end{equation}
\begin{equation}
\Gamma(n^3 P_0 \to \gamma \gamma) = 
        \frac{27\alpha_e^2 Q^4 |R'_{nP}(0)|^2}{m_c^4} 
        \left[1+\frac{1.8\alpha_s}{\pi}\right]
\end{equation}
\begin{equation}
\Gamma(n^3 P_2 \to \gamma \gamma) = 
      \frac{36\alpha_e^2 Q^4 |R'_{nP}(0)|^2}{5 m_c^4} 
      \left[1-\frac{16}{3} \frac{\alpha_s}{\pi}\right]
\end{equation}
where  $\alpha_e=1/137$  is the fine structure constant, $Q = 2/3$ is the quark charge and $R'(0)$ represents the first derivative of the radial wavefunction at $r = 0$. The results obtained are presented in Table~\ref{tab:digamma}.

\begin{table}[htbp] % The asterisk makes it span both columns
\centering
\small
\caption{Two-photon decay widths in keV.}
\label{tab:digamma}
\begin{tabular}{ lccccc }
\toprule
State    & This Work       & \cite{Cui} & \cite{Kher} & \cite{Soni} & PDG \cite{PDG}     \\
\midrule
$1^1S_0$ & 8.247 & 5.000 & 10.351      & 7.231       & $5.06 \pm 0.41$ \\
$2^1S_0$ & 3.494 & 3.000 & 4.501       & 5.507       & $2.09 \pm 1.19$ \\
$3^1S_0$ & 2.524 & 4.700 & 3.821       & 4.971       &                 \\
$4^1S_0$ & 2.070 & 3.500 & 3.582       & 4.688       &                 \\
\midrule
$1^3P_0$ & 2.240  &       & 1.973       & 8.982       & $2.16 \pm 0.20$ \\
$2^3P_0$ & 2.697  &       & 2.299       & 9.111       &                 \\
$3^3P_0$ & 2.892  &       & 2.714       & 9.104       &                 \\
$4^3P_0$ & 3.011  &       &             & 9.076       &                 \\
\midrule
$1^3P_2$ & 0.117  & 0.120 & 0.526       & 1.069       & $0.57 \pm 0.04$ \\
$2^3P_2$ & 0.139  & 0.160 & 0.613       & 1.085       &                 \\
$3^3P_2$ & 0.148  & 0.230 & 0.724       & 1.084       &                 \\
$4^3P_2$ & 0.153  & 0.240 &             & 1.080       &                 \\
\bottomrule
\end{tabular}
\end{table}
\begin{table}[htbp] % The asterisk makes it span both columns
\centering
\small
\caption{Two-gluon decay widths in MeV.}
\label{tab:digluon}
\begin{tabular}{ lcccc }
\toprule
State    & This Work & \cite{Kher} & \cite{Soni} & PDG \cite{PDG}  \\
\midrule
$1^1S_0$ & 22.309   & 24.249 & 35.909  & $28.6 \pm 2.2$  \\
$2^1S_0$ &  9.452   & 10.545 & 27.345  & $14.0 \pm 7.0$  \\
$3^1S_0$ &  6.827   &  8.952 & 24.683  &                 \\
$4^1S_0$ &  5.600   &  8.392 & 23.281  &                 \\
\midrule
$1^3P_0$ & 26.428   & 4.621  & 37.919  & $10.0 \pm 0.6$  \\
$2^3P_0$ & 31.803   & 5.386  & 38.462  &                 \\
$3^3P_0$ & 34.118   & 6.357  & 38.433  &                 \\
$4^3P_0$ & 35.524   &        & 38.315  &                 \\
\midrule
$1^3P_2$ & 2.083    & 1.232  & 3.974   & $1.97 \pm 0.11$ \\
$2^3P_2$ & 2.472    & 1.436  & 4.034   &                 \\
$3^3P_2$ & 2.632    & 1.695  & 4.028   &                 \\
$4^3P_2$ & 2.713    &        & 4.016   &                 \\
\bottomrule
\end{tabular}
\end{table}

The investigation of decay widths of charmonium states into two gluons is more complicated than two photon decays, since gluonic decays can only be observed indirectly from relatively stable hadronic final states. In the nonrelativistic limit, digluon decay width of a pseudoscalar meson is evaluated by~\cite{Lansberg} the following equations and results are listed in Table~\ref{tab:digluon}

\begin{equation}
\Gamma(n^1 S_0 \to gg) = 
      \frac{2\alpha_s^2 |R_{nsP}(0)|^2}{3m_c^2}
      \left( 1 + \frac{4.8\alpha_s}{\pi} \right)
\end{equation}

\begin{equation}
\Gamma(n^3 P_0 \to gg) = 
      \frac{6\alpha_s^2 |R'_{nP}(0)|^2}{m_c^4}
      \left( 1 + \frac{9.5\alpha_s}{\pi} \right)
\end{equation}

\begin{equation}
\Gamma(n^3 P_2 \to gg) = 
      \frac{4\alpha_s^2 |R'_{nP}(0)|^2}{5m_c^4} 
      \left( 1 - \frac{2.2\alpha_s}{\pi} \right)
\end{equation}

Vector mesons ($^3S_1$ and $^3D_1$) can decay into lepton pairs as well, since they have quantum numbers $1^{--}$. Dilepton decay width of the vector meson is given by~\cite{Royen} the following formulae and the calculated values are given in Table~\ref{tab:dilepton}.
\begin{equation}
\Gamma(n^3 S_1 \to e^+e^-) = \frac{4 \pi Q^2 \alpha^2 f_V^2}{3M_{nS}^2} \approx \frac{4 Q^2 \alpha^2 |R_{nS}(0)|^2}{M_{nS}^2} \left( 1 - \frac{16\alpha_s}{3\pi} \right) 
\end{equation}
\begin{equation}
\Gamma(n^3 D_1 \to e^+e^-) = \frac{25 Q^2 \alpha^2 |R''_{nD}(0)|^2}{2 m_Q^4 M_{nD}^2} \left( 1 - \frac{16\alpha_s}{3\pi} \right)
\end{equation}
\begin{table}[htbp] % The asterisk makes it span both columns
\centering
\small
\caption{Two-lepton decay widths in keV.}
\label{tab:dilepton}
\begin{tabular}{ lccccc }
\toprule
State    & This Work      & \cite{Cui} & \cite{Kher} & \cite{Soni} & PDG \cite{PDG}  \\
\midrule
$1^1S_1$ & 4.163 & 7.700 & 3.623       & 2.925       & $5.53 \pm 0.03$  \\
$2^1S_1$ & 1.623 & 3.100 & 1.085       & 1.533       & $2.27 \pm 0.07$  \\
$3^1S_1$ & 1.031 & 1.900 & 0.748       & 1.091       & $0.86 \pm 0.15$  \\
$4^1S_1$ & 0.756 & 1.300 & 0.599       & 0.856       & $0.35 \pm 0.14$  \\
\midrule
$1^3D_1$ & 0.121 & 0.147 & 0.113       &             & $0.261 \pm 0.021$\\
$2^3D_1$ & 0.359 & 0.150 & 0.166       &             & $0.476 \pm 0.238$\\
$3^3D_1$ & 0.744 & 0.126 & 0.211       &             &                  \\
$4^3D_1$ & 0.122 & 0.123 &             &             &                  \\
\bottomrule
\end{tabular}
\end{table}

\subsection{Electromagnetic Transition Widths}
The study of radiative transitions in heavy charmonium systems is essential for connecting charmonium states with different quantum numbers. These transitions may also improve theoretical models using experimental data. Radiative transitions can be determined by the electromagnetic current matrix elements that connect the initial $i$ and final $f$ states of the charmonium system. The electric dipole (E1) and magnetic dipole (M1) transitions are leading order transition amplitudes~\cite{Ding, Lu, Guo}. E1  radiative partial width is evaluated using
\begin{equation}
\Gamma_{\text{E1}}(n\, ^{2S+1}L_J \to n'\, ^{2S'+1}L'_{J'} + \gamma) =  \frac{4}{3} Q^2 \alpha C_{fi} \delta_{SS'} |\langle f|r|i \rangle|^2 
  E_\gamma^3 \frac{E_f}{M_i}
\end{equation}
%and
%\begin{equation}
%\Gamma_{\text{M1}}(n\, ^{2S+1}L_J \to n'\, ^{2S'+1}L'_{J'} + \gamma) =  \frac{4}{3} \frac{2J'+1}{2L+1} \delta_{LL'} \delta_{S, S' \pm 1} Q^2 \frac{\alpha}{m_c^2} |\langle f|i \rangle|^2
% E_\gamma^3 \frac{E_f}{M_i}
%\end{equation}
% 
where $E_\gamma$ is the photon energy in the rest frame of the initial state and $E_f$ is the energy of the final state. The matrix element
\begin{equation}
\langle f | r | i \rangle = \int_{0}^{\infty} r^2 dr \ R_{n', L'}(r) \ r \ R_{n, L}(r)
\end{equation}
involves the initial and final radial wavefunctions, $C_{fi}$ is the angular momentum matrix given by
\begin{equation}
C_{fi} = \max(L, L') (2J' + 1) \begin{Bmatrix} L' & J' & S \\ J & L & 1 \end{Bmatrix}^2
\end{equation}
Note that the primed symbols represent the final state quantum numbers.
We present only the result of E1 rates in Table~\ref{tab:E1transition} since M1 rates are considerably smaller.

\begin{table}[htbp] % The asterisk makes it span both columns
\centering
\small
\caption{E1 transition widths of some $c \bar c$  states in keV. The radiative decays $X(3872)\to \psi(2S)\gamma$ and $X(3872)\to J/\psi(1S)\gamma$ are indicated by * symbol.}
\label{tab:E1transition}
\begin{tabular}{ lrrrr }
\toprule
Transition    & This Work & \cite{Kher} & \cite{Soni} & PDG \cite{PDG}  \\
\midrule
$2^3S_1 \to 1^3P_0$ &    9.80  &  11.93  &   21.86 &  $27.9 \pm 1.7$ \\
$2^3S_1 \to 1^3P_1$ &   22.40  &  10.39  &   43.29 &  $27.9 \pm 1.7$ \\
$2^3S_1 \to 1^3P_2$ &   46.57  &   7.07  &   62.31 &  $26.8 \pm 1.6$ \\
$2^1S_0 \to 1^1P_1$ &   59.55  &   7.94  &   36.20 &                 \\
$1^3P_0 \to 1^3S_1$ &  410.34  & 118.29  &  112.03 &  $148 \pm 15$   \\
$1^3P_1 \to 1^3S_1$ &  439.81  & 189.86  &  146.32 &  $288 \pm 18$   \\
$1^3P_2 \to 1^3S_1$ &  416.52  & 233.85  &  157.23 &  $384 \pm 22$   \\
$1^1P_1 \to 1^1S_0$ &  481.85  & 357.83  &  247.97 &  $450 \pm 171$  \\
$2^3P_0 \to 2^3S_1$ &  261.48  & 102.23  &   70.40 &                 \\
$2^3P_1 \to 2^3S_1$*&  278.65  & 206.87  &  102.67 &                 \\
$2^3P_2 \to 2^3S_1$ &  268.35  & 281.93  &  116.33 &                 \\
$2^1P_1 \to 2^1S_0$ &  283.34  & 343.55  &  163.65 &                 \\
$2^3P_0 \to 1^3S_1$ &  119.70  &         &  173.32 &                 \\
$2^3P_1 \to 1^3S_1$*&  132.69  &         &  210.96 & $11.9 \pm 5.2$  \\
$2^3P_2 \to 1^3S_1$ &  111.76  &         &  227.92 &                 \\
$2^1P_1 \to 1^1S_0$ &  133.77  &         &  329.38 &                 \\
$2^3P_2 \to 1^3D_2$ &   10.72  &   5.49  &         &                 \\
$2^3P_2 \to 1^3D_1$ &    0.82  &   0.41  &         &                 \\
$1^3D_1 \to 1^3P_0$ &  190.54  & 343.87  &  161.50 &  $188 \pm 18$   \\
$1^3D_1 \to 1^3P_1$ &  126.05  & 139.52  &   93.78 &  $67.7 \pm 6.7$    \\
$1^3D_1 \to 1^3P_2$ &    9.24  &   6.45  &    5.72 &  $ < 18 $       \\
$1^3D_2 \to 1^3P_1$ &  225.64  &  89.18  &  165.18 &                 \\
$1^3D_2 \to 1^3P_2$ &   83.09  &  62.34  &   50.32 &                 \\
\bottomrule
\end{tabular}
\end{table}

%%%%%%%%%%%%%%%%%%%%%%%%%%%%%%%%%%%%%%%%%%%%%%%%%%%%%%%%%%%%%%%%%%%%%%%%%%%%%%
\section{Radiative Decays of X(3872) Meson}
\label{sec:x3872}

The discovery of the $X(3872)$ is first reported by the Belle Collaboration \cite{Belle} in the $J/\psi \pi^+ \pi^-$ invariant mass spectra in the process $ e^+ e^- \to \Upsilon(4S) \to B^+ B^-$ and $B^\pm \to K^\pm J/\psi \pi^+ \pi^-$ in 2003.
Following its discovery, one of the primary experimental goal was to determine its quantum numbers. In 2013, the LHCb Collaboration performed an angular correlation study of $B^+ \to K^+ X(3872)$ decays and successfully assign the quantum numbers $J^{PC} = 1^{++}$~\cite{LHCbJPC}. More recently, in 2024, the LHCb Collaboration extended the study of its radiative decay modes using proton-proton collision data~\cite{LHCbratio} where they reported the observation of the $X(3872) \to \psi(2S)\gamma$ decays with relatively large number of data. A significant result of this study is the measurement of the ratio of partial widths
\begin{equation}
    R_{\psi\gamma} = \frac{\mathcal{B}(X(3872) \to \psi(2S)\gamma)}{\mathcal{B}(X(3872) \to J/\psi\gamma)} = 1.67 \pm 0.25
\end{equation}
where the error represents the total combined uncertainty. This measurement can provide critical input for theoretical models to distinguish whether the $X(3872)$ is a standard charmonium state, a $D\bar{D}^*$ molecule, or a more complex exotic state.

Using our values summarized in Table~\ref{tab:E1transition}, the predicted ratio of transition widths is found to be $R_{\psi\gamma} = 2.10$. Although the calculated mass, $M(X(3872)) = 3968$~MeV in Table~\ref{tab:fullmassspectra}, is about $2.5\%$ higher than the observed value, the ratio $R_{\psi\gamma}$ remains consistent with the results obtained by LHCb using Run1 and Run2 data.
Even though the calculated $R_{\psi\gamma}$ is within $2 \sigma$ of the observed value, as can be seen in Table \ref{tab:E1transition}, the calculated value of transition width of $2^3P_1 \to 1^3S_1$ is one order of magnitude larger than the experimental value. 
Note that in Ref~\cite{Grinstein:2024}, this ratio is estimated to be much smaller than the experimental value. In Ref~\cite{Cincioglu:2016fkm}, it was shown that a larger molecular component implies a smaller ratio. If a pure charmonium state overestimates the ratio, including a molecular component will lower the ratio, and hence a large radiative decay (as obtained in the present work) can be interpreted as the presence of a molecular component. The determination of the exact fraction of molecular component is beyond the scope of this study.
%This might be interpreted as the existence of a significant molecular component in $X(3872)$ as this decay is expected to be suppressed in a molecular $X(3872)$ due to the very large size of the molecule.
%

In Ref~\cite{Colangelo:2025uhs}, the same ratio has been studied within a heavy quark effective field theory framework where $X(3827)$ is treated as a member of a heavy quark quadruplet. The value obtained $R_{\psi \gamma} = 1.7 \pm 0.3$ is consistent with the experimental result. Note that in \cite{Colangelo:2025uhs}, the considered quadruplet can be charmonia with $L=1$, or can also be a quadruplet of tetraquarks, or even a mixture of them. 
For this reason, the interpretation of the better agreement of their result with the experimental data in terms of the structure of $X(3872)$ is nontrivial. 
%Hence, their better agreement with the experimental result does not give a deeper understanding to the internal structure of $X(3872)$.

%%%%%%%%%%%%%%%%%%%%%%%%%%%%%%%%%%%%%%%%%%%%%%%%%%%%%%%%%%%%%%%%%%%%%%%%%%%%%%
\section{Summary and Conclusion}
\label{sec:theend}

In this study, we investigated charmonium states within the framework of a non-relativistic potential model by introducing a novel confinement function. The potential parameters and the specific form of the confinement part of the potential function were extracted using a stochastic (Monte Carlo) approach. This phenomenological stochastic approach provides a useful baseline for deriving the confinement part of the interaction potential function and can be further refined and extended in future work.

The optimized potential parameters are subsequently used to determine the mass spectra and predict the decay properties of the charmonium system. Our results are mostly in good agreement with the experimental values reported by PDG. Moreover, the calculated ratio of transition widths of $X(3872)$, $R_{\psi \gamma}$, is found to be consistent with the recent observation by the LHCb Collaboration. Nevertheless, the observed discrepancy between the actual values of the radiative decay widths with the observed values can be interpreted as a sign of a significant molecular component in $X(3872)$.

We also compare our results with some other theoretical studies based on various potential models. While these models and our study generally results in an accurate description of the charmonium mass spectra in agreement with PDG data, their predictions for decay properties exhibit quite significant deviations from each other and from experimental observations.

The synthetic benchmark presented in Appendix shows that the unsorted-slope sampling can have difficulty recovering strongly screened asymptotic behavior because such configurations are strongly suppressed by the proposal distribution. A sorted-slope modification can recover the plateau in the benchmark case, but this modification introduces its own sampling bias and is therefore not used in the main charmonium analysis.

%%%%%%%%%%%%%%%%%%%%%%%%%%%%%%%%%%%%%%%%%%%%%%%%%%%%%%%%%%%%%%%%%%%%%%%%%%%%%%
\section*{Appendix: Benchmark Test on Synthetic Data}
To test the performance of proposed stochastic potential extraction algorithm on saturating shapes, we perform an inverse reconstruction benchmark using synthetic mock.

We start with a simplified bound-state radial Hamiltonian in dimensionless units where $\hbar^2/m_r = 1$:
\begin{equation}
\label{eqn:newhamiltonian}
  H = -\frac{1}{2}\frac{d^2}{dr^2}  + V_{\text{true}}(r)
\end{equation}
Here the target potential $V_{\text{true}}(r)$ includes a screened confinement term:
\begin{equation}
\label{eqn:screening}
  V_{\text{true}}(r) = -\frac{4\alpha_s}{3r} + \frac{\sigma}{\mu} \left(1-e^{-\mu r} \right)
\end{equation}
For the sake of simplicity, the spin-dependent potential terms are omitted ($L=S=0$) in this benchmark. This setup allows us to generate high-precision numerical reference eigenvalues, which we call $\{E_{nS}^{\text{exact}}\}$, and test whether the inverse stochastic sampler can accurately recover $V_{\text{true}}(r)$\footnote{The C++/ROOT implementation used for this simulation is available at\\ \url{https://github.com/abingul/Stocastic_QCD_Potential}}.

\begin{figure}[htbp]
\centering
\includegraphics[width=0.7\textwidth]{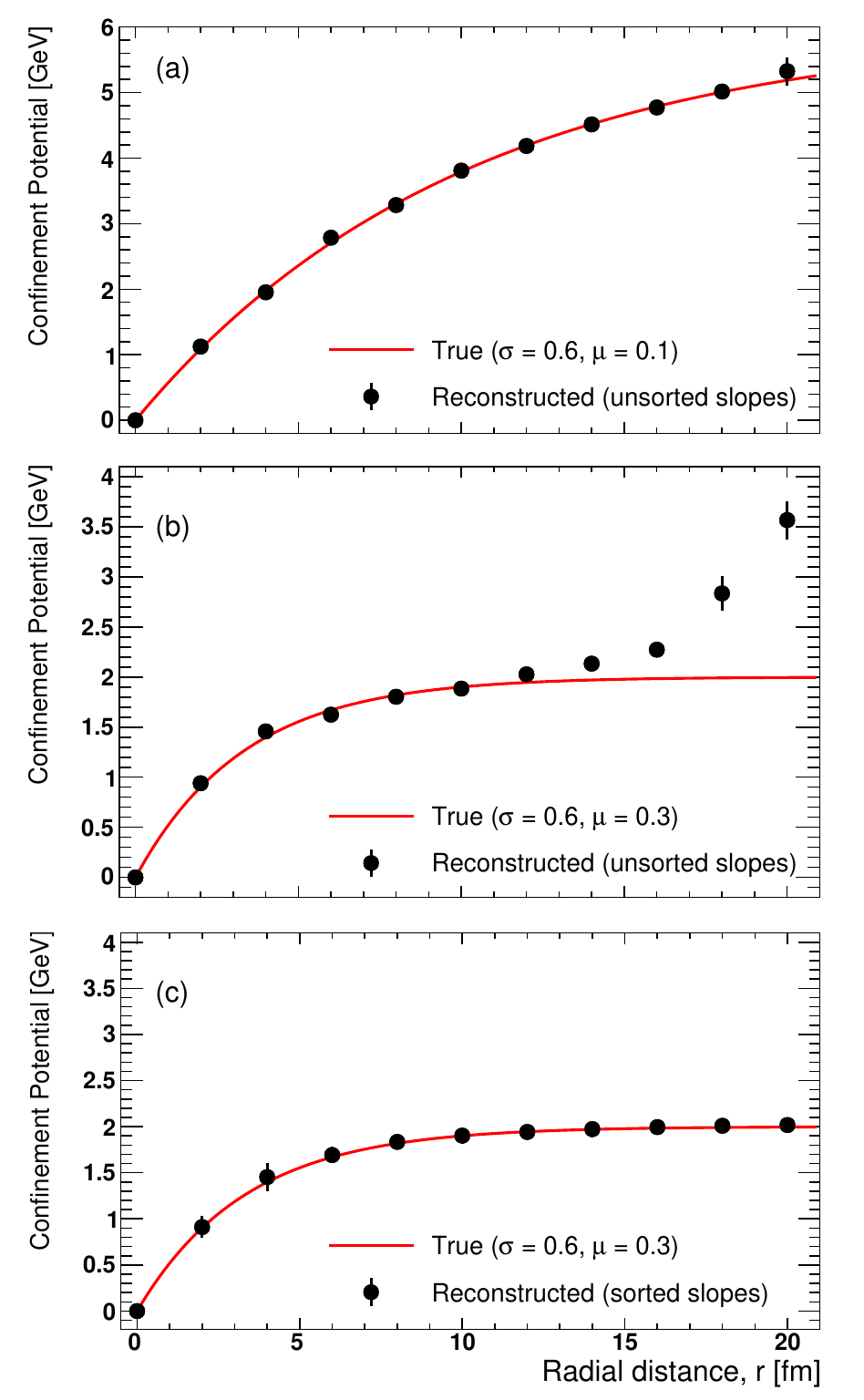}
\caption{
(Color online) Comparison between the benchmark true screening potentials (solid line) and reconstructed stochastic potentials (data points) across $N = 11$ radial points. Reconstructed values represent ensemble averages over $N_{\text{config}} = 6$ independent configurations initiated with different random seeds, where error bars denote one standard deviation. (a) Mild screening ($\sigma = 0.6$, $\mu = 0.1$) using standard unsorted slope increments. (b) Stronger screening ($\sigma = 0.6$, $\mu = 0.3$) using unsorted slopes, exhibiting a large-distance divergence. (c) Stronger screening ($\sigma = 0.6$, $\mu = 0.3$) using monotonically sorted slopes ($c_{j+1} \le c_j$), successfully capturing the flat asymptotic plateau.}
\label{fig:app}
\end{figure}

Figure~\ref{fig:app}a shows a comparison between the benchmark screening potential and the reconstructed stochastic confinement potential using input parameters $\alpha_s = 0.4$, $\sigma = 0.6\text{ GeV/fm}$, and $\mu = 0.1\text{ fm}^{-1}$. Candidate slope increments are drawn independently from a uniform distribution $c_j \sim U[0, 1/2]$. As summarized in Table~\ref{tab:mu01}, the stochastic inverse reconstruction successfully reproduces the screening profile and the exact eigenvalues of the target energy within the statistical uncertainty bands. 

Conversely, when the screening parameter is increased to $\mu = 0.3\text{ fm}^{-1}$ (for $\alpha_s = 0.4$ and $\sigma = 0.6\text{ GeV/fm}$), the standard algorithm fails to recover the true potential at larger interquark distances, exhibiting a divergent behavior as shown in Figure~\ref{fig:app}b. This failure confirms that the algorithm cannot sample the flat asymptotic plateau when slopes are taken from an unconditioned uniform distribution. As a result, even with millions of iterations, the standard unsorted stochastic approach described in Section~\ref{sec:stochastic} is strongly suppressed from recovering such saturating profiles because of this phase-space limitation.

However, when the algorithm is modified to enforce decreasing slope increments (that is, if the slopes are sorted in decreasing order so that $c_{j+1} \le c_j$ where $c_j \sim U[0, 1/2]$), the stochastic algorithm successfully recovers the true screened potential profile for all radial regions, including the asymptotic plateau. This is illustrated in Figure~\ref{fig:app}c and the evaluated eigenvalues are listed in Table~\ref{tab:mu03}. Consequently, incorporating ordered slope bounds removes the large-distance divergence and restores the algorithm's capability to capture the saturating screening shape.

To understand the origin of this issue, in Tables~\ref{tab:mu01} and \ref{tab:mu03}, we also provided the expectation value $\langle r \rangle$ obtained using the reduced wavefunction ($u(r)$).
Typically, $u(r)$ is exponentially suppressed for $r > \langle r \rangle$. The eigenvalue equation is insensitive to the values of the potential in regions where $u(r) \sim 0$, i.e. in this case with states upto $n=8$, the eigenvalue equation can only probe the potential in the regions $r < \langle r \rangle \lesssim 12$ in natural units. The unsorted algorithm is already biased against asymptotically flat potential. 
Combined with the insensitivity to the value of the potential in the region $r>12$ leads to a prediction that increases with $r$. Note that, although sorting the slopes $c_j$ fixes the bias against asymptotically flat potentials, it introduces a bias in favor of them.

We emphasize that the main charmonium calculations presented in Section~\ref{sec:method} rely on the unsorted slope algorithm. The sorted-slope modification is introduced only in this Appendix to demonstrate how the algorithmic divergence at large interquark distances can be resolved for strongly screened interactions.

\begin{table}[htbp] % The asterisk makes it span both columns
\centering
\scriptsize
\caption{Comparison of reference (exact) and extracted energy eigenvalues for mild screening ($\sigma = 0.6$, $\mu = 0.1$), corresponding to Figure~\ref{fig:app}a. Standard deviations $\sigma_{E_n}$ are evaluated across $N_{\text{config}} = 6$ independent stochastic realizations. The expectation value $\langle r \rangle$ (in natural units) for each state is also provided.}
\label{tab:mu01}
\begin{tabular}{ llcccccccc }
\toprule
         & $n$                    & 1     & 2     & 3     & 4     & 5     & 6     & 7     & 8     \\
\midrule
         &$E_{n}^{\text{exact}}$ & 0.734 & 1.762 & 2.457 & 3.001 & 3.450 & 3.831 & 4.159 & 4.445 \\
\midrule
Unsorted & $E_{n}^{\text{mean}}$  & 0.734 & 1.754 & 2.459 & 3.021 & 3.447 & 3.831 & 4.161 & 4.438 \\
slopes   & $\sigma_{E_n}$         & 0.028 & 0.026 & 0.014 & 0.016 & 0.023 & 0.024 & 0.026 & 0.011 \\
\midrule
         & $\left\langle r \right\rangle$   & 1.259 & 2.614 & 3.808 & 4.952 & 6.087 & 7.238 & 8.421 & 9.652 \\
\bottomrule
\end{tabular}
\end{table}
\begin{table}[htbp] % The asterisk makes it span both columns
\centering
\scriptsize
\caption{Comparison of reference (exact) and extracted energy eigenvalues for stronger screening ($\sigma = 0.6$, $\mu = 0.3$), corresponding to Figure~\ref{fig:app}b, \textit{unsorted} potential slope increments and
Figure~\ref{fig:app}c, \textit{sorted} potential slope increments. Standard deviations $\sigma_{E_n}$ are evaluated across $N_{\text{config}} = 6$ independent stochastic realizations. The expectation value $\langle r \rangle$ (in natural units) for each state is also provided.}
\label{tab:mu03}
\begin{tabular}{ llcccccccc }
\toprule
         & $n$                    & 1     & 2     & 3     & 4     & 5     & 6     & 7     & 8     \\
\midrule
         & $E_{n}^{\text{exact}}$ & 0.633 & 1.389 & 1.746 & 1.917 & 2.241 & 2.409 & 2.604 & 2.827 \\
\midrule
Unsorted & $E_{n}^{\text{mean}}$  & 0.617 & 1.382 & 1.721 & 1.957 & 2.194 & 2.393 & 2.612 & 2.846 \\
slopes   & $\sigma_{E_n}$         & 0.029 & 0.031 & 0.022 & 0.022 & 0.021 & 0.019 & 0.019 & 0.013 \\
\midrule
Sorted   & $E_{n}^{\text{mean}}$  & 0.594 & 1.393 & 1.752 & 1.919 & 2.250 & 2.417 & 2.612 & 2.834 \\
slopes   & $\sigma_{E_n}$         & 0.035 & 0.040 & 0.023 & 0.031 & 0.030 & 0.031 & 0.030 & 0.029 \\
\midrule
         & $\left\langle r \right\rangle$ & 1.378 & 3.305 & 5.787 & 9.703 &10.866 &11.028 &11.246 & 11.553 \\
\bottomrule
\end{tabular}
\end{table}

\label{sec:Appendix}

\newpage

%%%%%%%%%%%%%%%%%%%%%%%%%%%%%%%%%%%%%%%%%%%%%%%%%%%%%%%%%%%%%%%%%%%%%%%%%%%%%%
\bigskip
\small
\noindent \textbf{Acknowledgements}\\
We are grateful to Tahmasib Aliyev, Soner Albayrak, Caner \"Unal, Do\u{g}a Veske, Seçkin Kürkçüoğlu and İsmail Turan for their suggestions and encouragement regarding the manuscript. A.B. thanks the Middle East Technical University Physics Department for their hospitality throughout this study.

%%%%%%%%%%%%%%%%%%%%%%%%%%%%%%%%%%%%%%%%%%%%%%%%%%%%%%%%%%%%%%%%%%%%%%%%%%%%%%

%\bibliographystyle{refs-style}
\bibliographystyle{unsrt} % <--- CHANGE THIS
\bibliography{refs}
\end{document}